\documentclass[oneside,10pt]{article}

\newtheorem{example}{Example}

\begin{document}

\title{Parallel Algorithms for DNA Probe Placement on Small Oligonucleotide Arrays}

\author{Drago\c{s} Trinc\u{a} and Sanguthevar Rajasekaran\\
Department of Computer Science and Engineering\\University of Connecticut\\Storrs, CT 06269, USA\\\{dtrinca,rajasek\}@engr.uconn.edu\\
}
\maketitle

\begin{abstract}
		Oligonucleotide arrays are used in a wide range of genomic analyses, such as gene expression profiling,
		comparative genomic hybridization, chromatin immunoprecipitation, SNP detection, etc.
		During fabrication, the sites of an oligonucleotide array are selectively exposed to light in order to activate oligonucleotides
		for further synthesis. Optical effects can cause unwanted illumination
		at masked sites that are adjacent to the sites intentionally exposed to light. This results in
		synthesis of unforeseen sequences in masked sites and compromises interpretation
		of experimental data. To reduce such uncertainty, one can exploit freedom in how
		probes are assigned to array sites. The {\it border length minimization problem} (BLMP) seeks a placement of probes that
		minimizes the sum of border lengths in all masks.
		In this paper, we propose two parallel algorithms for the BLMP.
		The proposed parallel algorithms have the local-search paradigm at their core, and are especially
		developed for the BLMP. The results reported show that, for small microarrays with at most $1156$ probes,
		the proposed parallel algorithms perform better than the best previous algorithms.
\end{abstract}


%

\section{Introduction}
	Oligonucleotide arrays, such as those produced by Affymetrix \cite{affymetrix1}, are used in a wide range of genomic analyses.
	As discussed in \cite{fodor1,mandoiu1}, during very large-scale immobilized polymer synthesis (VLSIPS) the sites of
	a DNA probe array are selectively exposed to light in order to activate oligonucleotides for further synthesis. The selective exposure
	is achieved by a sequence of masks, with each mask consisting of nontransparent and
	transparent regions corresponding to the masked and exposed array sites.
	Optical effects (diffraction, reflections, etc.) can cause unwanted illumination
	at masked sites that are adjacent to the sites intentionally exposed to light
	- i.e., at the border sites of transparent regions in the mask. This results in
	synthesis of unforeseen sequences in masked sites and compromises interpretation
	of experimental data. To reduce such uncertainty, one can exploit freedom in how
	probes are assigned to array sites. The {\it border length minimization problem} (BLMP) seeks a placement of probes that
	minimizes the sum of border lengths in all masks. In this paper, we consider the synchronous version of the BLMP, which can be formulated
	as follows.
\begin{description}
\item[BLMP: ] Given a set ${\it SP}$ consisting of ${\it dim}^{2}$ DNA sequences (called {\it probes}) of the same length, place the probes
	in ${\it SP}$ on a ${\it dim}\times{\it dim}$ microarray in such a way that the sum of the Hamming distances between every two neighbors
	on the microarray is minimized. (Two probes on the microarray are said to be {\it neighbors} if: (1) they are adjacent and (2) they are
	on the same row or column of the microarray. Thus, for a ${\it dim}\times{\it dim}$ microarray, there are ${\it dim}*({\it dim}-1)*2$
	distinct pairs of neighbors.)
\end{description}
	The BLMP is important not only for arrays fabricated by Affymetrix, but also for any other in-situ synthesis scheme, such as the
	highly-efficient micromirror arrays \cite{sg1,invitrogen1,febit1}, or the membrane-based microarrays \cite{SABiosciences1}.

	Previous work on the BLMP consists of the following heuristics: the TSP+1-Threading algorithm proposed in \cite{hannenhalli1}, the epitaxial
	algorithm proposed in \cite{mandoiu1}, the row-epitaxial algorithm proposed in \cite{mandoiu2}, and the recursive partitioning
	algorithm proposed in \cite{mandoiu3}. We detail these heuristics as follows.
\begin{enumerate}
\item The TSP+1-Threading heuristic proposed in \cite{hannenhalli1} consists of two steps:
	(1) arrange the probes in a TSP tour by applying an approximation algorithm for the TSP, and then (2) place the sequence obtained in
	the first step on the microarray using the 1-Threading model, as described in \cite{hannenhalli1}.
\item The epitaxial heuristic proposed in \cite{mandoiu1} works as follows. Initially, a random probe in ${\it SP}$ is placed
	on a random location on the microarray, and then removed from ${\it SP}$. Then, as long as there is at least one probe in ${\it SP}$,
	do the following: randomly select an empty location on the microarray out of those with a maximum number of neighbors, and on that location
	place one of the probes in ${\it SP}$ that minimizes the sum of the Hamming distances between that probe and the current neighbors
	of the location (such a probe is randomly chosen if there are multiple probes that give the same minimum). Once the probe is placed,
	it is removed from ${\it SP}$.
\item The row-epitaxial heuristic proposed in \cite{mandoiu2} is similar to the epitaxial heuristic. The main difference
	consists of the fact that instead of placing the probes one by one, it re-shuffles an already existing pre-optimized placement.
\item The recursive partitioning heuristic proposed in \cite{mandoiu3} partitions the set of probes into subsets of
	the same size, and then places the probes in each subset on the corresponding submicroarray.
\end{enumerate}
	
	In this paper, we propose two parallel algorithms for the BLMP that are shown to give better results than the previous algorithms
	for small microarrays with up to $1156$ probes. Such algorithms are especially useful for companies like:
\begin{enumerate}
\item SABiosciences \cite{SABiosciences1}, which fabricates custom microarrays with just 440 probes;
\item Febit \cite{febit1}, which fabricates small microarrays with just a few thousand probes.
\end{enumerate}
\section{Results and Discussion}
	In this section, we propose several algorithms for the BLMP. The algorithms that we propose are based on the local-search
	paradigm \cite{ls1}, but are more involved, and especially designed for the BLMP.
\subsection{A Local-Search-based Sequential Algorithm}
	In this section, we propose a local-search-based sequential algorithm for the BLMP, called LS. It is given in Fig. \ref{fig:LS}.
	It works as follows. It takes as input the set of probes, a time limit $T$, and a probability parameter ${\it pr}$.
	Initially, the probes are randomly placed on the microarray.
	As long as the time limit was not exceeded yet, the algorithm randomly selects two locations, say $(l_{1},c_{1})$ and $(l_{2},c_{2})$.
	If swapping the probes currently on these two locations leads to a decreasing in the cost, then they are swapped and the cost is updated.
	Otherwise, they are swapped only with a certain probability. When the time limit is exceeded, the best microarray
	configuration found during the algorithm is returned.
\subsection{A Local-Search-based Parallel Algorithm}
	The local-search-based idea can be easily parallelized, as shown in Fig. \ref{fig:LS-Par}. The parallel variant is called LS-Par.
	In the first step, processor $P_{1}$ places the probes randomly on its microarray, and then sends its microarray configuration to all
	the other processors. So, when reaching line 3, each processor has the same configuration. At the end of each iteration through the WHILE
	loop that starts at line 11, the processors synchronize with each other. Let $P_{\it source}$ be randomly selected out of those processors
	with a minimum ${\it COST}$. All the other processors update their variables with the corresponding variables from $P_{\it source}$.
	So, in conclusion, all the processors enter and exit every iteration through the WHILE loop with the same microarray configuration.
	The final result is returned by one of the processors, say $P_{1}$.
\subsection{ALG1}
	Let $P_{1},P_{2},\ldots,P_{k}$ be the available processors.
	In this section, we propose a parallel algorithm for the BLMP, called ALG1, which is given in Fig. \ref{fig:ALG1}. The code shown
	in Fig. \ref{fig:ALG1} is executed by each of the $k$ processors separately. At each step, the processors synchronize with each other
	and (possibly) exchange some data. ALG1 incorporates the local-search idea we have seen in LS and LS-Par, but is more complicated, and
	especially developed for the BLMP.
	
	The details are as follows. Each of the processors takes as input the same parameters, namely: a set of probes ${\it SP}$, a time limit
	$T$, and positive integers ${\it probelength}$, ${\it MaxTrials1}$, ${\it MaxTrials2}$, ${\it MaxCost1}$, ${\it MaxCost2}$, ${\it winlength}1$,
	and ${\it winlength2}$.
	The goal is to find a placement of the probes in ${\it SP}$ on the microarray as close as possible to the optimal.
	
	The algorithm proceeds as follows. First, processor $P_{1}$ places the probes in ${\it SP}$ on the microarray, and then sends its
	microarray configuration to all the other processors. So, initially, all the processors have the same placement on the microarray.
	Then, each of the processors copies its microarray to ${\it bestmicroarray}$, where ${\it bestmicroarray}$ is the microarray
	that keeps the best configuration found during the algorithm. So, initially, each processor has the same ${\it bestmicroarray}$.
	In lines 6-9, each of the processors computes the ${\it COST}$ of the initial microarray configuration, so each of the processors
	has the same ${\it COST}$. Also, ${\it bestCOST}$ is the cost corresponding to the ${\it bestmicroarray}$ configuration.
	In ${\it average}$, each of the processors keeps the current average Hamming distance between any two neighbors on the current microarray
	configuration. (Note that ${\it dim}*({\it dim}-1)*2$ is the total number of pairs of neighbors on a microarray with ${\it dim}^{2}$
	locations.)
	
	The basic step of ALG1 starts at line 15 and ends at line 66. It is repeated until the time taken by the algorithm exceeds $T$.
	Each of the processors repeats this basic step the same number of times. If one of the processors exits the WHILE loop that starts
	at line 14, then it will notify the other processors, so that the other processors will not wait for the synchronizations that start
	at lines 36 and 47.

	During each basic step, each of the processors tries to find a pair of locations on the current microarray, say
	$(l_{1},c_{1})$ and $(l_{2},c_{2})$, with the property that swapping the probes that are currently on those locations, namely
	${\it microarray}[l_{1},c_{1}]$ and ${\it microarray}[l_{2},c_{2}]$, will lead to a decreasing in the cost (or to a cost equal to the
	current cost). The pair of locations
	that are examined during each basic step are randomly generated, and depend on the current processor, since each processor has its own
	random number generator. The IF statement that starts at line 32 tries to see if swapping the probes on the chosen locations leads to
	an increasing in the
	cost (or to a cost equal to the current cost). If yes, then the probes are swapped, the ${\it OK}$ variable is set to $1$ (meaning that
	the current basic step is finished),
	and the ${\it COST}$ variable is updated accordingly. After the IF statement that starts at line 32, the processors synchronize with
	each other.
	If at least one of them, say $P_{\it source}$, has ${\it OK}=1$, then that means that at least one of them has succeeded in finding a pair
	of locations that leads to a decreasing in the ${\it COST}$ (or to a cost equal to the current ${\it COST}$). If so, then all the
	processors (including those that have ${\it OK}=1$)
	update their microarray configuration with the microarray from $P_{\it source}$. In such a case, all the processors will have ${\it OK}=1$
	after the synchronization that starts at line 36, and thus, all of them will exit the WHILE loop that starts at line 17.

	If none of the processors has ${\it OK}=1$, then all the processors will have ${\it OK}=0$ after the synchronization that starts at line 36,
	and thus, all of them will enter the IF statement that starts at line 40, and then synchronize with each other at line 47. If, when
	reaching line 47, at least one of them, say $P_{\it source}$, has ${\it OK}=1$, then that means that all the other processors will update
	their {\it microarray} and ${\it COST}$ with the corresponding variables from $P_{\it source}$, and then set their ${\it OK}$ variable
	to $1$. (At line 42, ${\it average}$ is multiplied by $8$ since the two chosen locations have at most $8$ neighbors in total.)
	
	So, in conclusion, all the processors exit the WHILE loop that starts at line 17 with the same ${\it nrt}$, meaning that each of the processors
	has tried the same number of pairs of locations before setting its ${\it OK}$ variable to $1$ (or reaching the ${\it MaxTrials2}$ limit
	at line 51). Having this said, it is clear that ${\it MaxTrials2}$ is meant to help the processors exit the WHILE loop that starts
	at line 17, and not let them run indefinitely in case that ${\it OK}=0$ at all the processors after each processor has tried at least
	${\it MaxTrials2}$ pairs of locations. Also, note that all the processors enter and exit the WHILE loop that starts at line 17
	with the same microarray configuration. This also implies that all of them will eventually exit the WHILE loop that starts at line 14
	with the same microarray configuration and the same ${\it bestmicroarray}$.
	
	In lines 53-66, each processor updates its corresponding parameters. The ${\it average}$ variable is updated according to the new
	${\it COST}$. In case that ${\it COST}<{\it bestCOST}$, then ${\it bestCOST}$ and ${\it bestmicroarray}$ are updated accordingly at each
	processor. If ${\it myub}=0$ and the WHILE loop that starts at line 17 was executed at least ${\it winlength1}$ times since
	the last update of ${\it myub}$, then ${\it myub}$ is set to ${\it MaxCost1}$. Otherwise, if ${\it myub}={\it MaxCost1}$ and the
	WHILE loop that starts at line 17 was executed at least ${\it winlength2}$ times since the last update of ${\it myub}$, then
	${\it myub}$ is set to $0$. So, in other words, when ${\it myub}={\it MaxCost1}$, the processors are allowed to shuffle more probes on the
	microarray (at line 42). This helps the algorithm to converge much faster to an approximate solution. The parameter ${\it MaxCost2}$ at line
	43 allows the processors to swap the probes as long as the overall cost of the resulting microarray configuration is under a certain
	threshold.
\begin{example}
	To see how ALG1 works, we give an example with three processors $P_{1}$, $P_{2}$, $P_{3}$, for a microarray of size $4\times{4}$.
	The input parameters are as follows:
\begin{itemize}
\item ${\it SP}$ has $16$ probes, each of length ${\it probelength}=5$;
\item $T$ is 10 seconds;
\item ${\it dim}=4$, meaning that the microarray is of size $4\times{4}$;
\item ${\it MaxTrials1}=2$;
\item ${\it MaxTrials2}=1000$;
\item ${\it MaxCost1}=10$;
\item ${\it MaxCost2}=10$;
\item ${\it winlength1}=70$;
\item ${\it winlength2}=20$.
\end{itemize}
	The probes in ${\it SP}$ are as given in Table \ref{table:Probes}. The algorithm works as follows.
\begin{description}
\item[INIT:] Processor $P_{1}$ randomly places the probes on its microarray. Without loss of generality, suppose
	that $P_{1}$ places the probes as shown in Fig. \ref{fig:InitialPlacement}. Processor $P_{1}$ sends its microarray configuration to
	$P_{2}$ and $P_{3}$. So, initially, all the processors have the same microarray configuration.
	In such a case, it can be seen that the initial cost is $85$ (so, ${\it bestCOST}$ is $85$ as well). This implies that initially,
	${\it average}=3.54$.
\item[Step 1:] At the beginning of this step, ${\it OK}=0$ and ${\it nrt}=0$. Also, ${\it sa}=0$, ${\it myub}=0$, and
	${\it average}=3.54$. ${\it COST}$ and ${\it bestCOST}$ are both $85$.
\begin{description}
\item[Substep 1:] Suppose that processor $P_{1}$ randomly selects locations $(1,1)$ and $(4,3)$. For these locations,
	${\it localcost}=16$, whereas ${\it newlocalcost}=17$. Processor $P_{2}$ randomly selects locations $(3,3)$ and $(1,1)$.
	For these locations, ${\it localcost}=22$ and ${\it newlocalcost}=20$. Processor $P_{3}$ randomly selects locations $(4,2)$ and $(1,4)$.
	For these locations, ${\it localcost}=14$ and ${\it newlocalcost}=15$. So, only processor $P_{2}$ enters the IF statement that starts
	at line 32. So, $P_{2}$ will reach the synchronization that starts at line $36$ with ${\it OK}=1$ and ${\it COST}=83$, and thus it will
	be the source processor. So, all the other processors exit the WHILE loop that starts at line 17.
\item[Update:] At the beginning of this phase, all the processors have ${\it COST}=83$. The ${\it average}$ variable becomes
	$3.45$, ${\it bestCOST}$ and ${\it bestmicroarray}$ are updated accordingly, ${\it sa}$ becomes $1$, and ${\it myub}$ remains $0$.
\end{description}
\item[Step 2:] At the beginning of this step, ${\it OK}=0$ and ${\it nrt}=0$. Also, ${\it sa}=1$, ${\it myub}=0$, and
	${\it average}=3.45$. ${\it COST}$ and ${\it bestCOST}$ are both $83$.
\begin{description}
\item[Substep 1:] Suppose that $P_{1}$ selects $(1,2)$ and $(4,1)$, $P_{2}$ selects $(1,2)$ and $(4,2)$, and $P_{3}$ selects
	$(3,1)$ and $(4,3)$. For $P_{1}$, ${\it localcost}=18$ and ${\it newlocalcost}=18$. For $P_{2}$, ${\it localcost}=22$ and
	${\it newlocalcost}=18$. For $P_{3}$, ${\it localcost}=19$, whereas ${\it newlocalcost}=24$. So, only $P_{1}$ and $P_{2}$ enter
	the IF statement that starts at line 32. Suppose that $P_{1}$ is randomly chosen to be the source processor. So, all the processors exit
	the WHILE loop that starts at line 17 (with ${\it COST}=83$).
\item[Update:] At this point, all the processors have ${\it COST}=83$. All the other variables remain unchanged, except for
	${\it sa}$, which becomes $2$.
\end{description}
\item[Step 3:] At the beginning of this step, ${\it OK}=0$ and ${\it nrt}=0$. Also, ${\it sa}=2$, ${\it myub}=0$, and
	${\it average}=3.45$. ${\it COST}$ and ${\it bestCOST}$ are both $83$.
\begin{description}
\item[Substep 1:] Suppose that $P_{1}$ selects $(1,4)$ and $(3,3)$, $P_{2}$ selects $(4,2)$ and $(1,2)$, and $P_{3}$ selects
	$(1,2)$ and $(2,4)$. For $P_{1}$, ${\it localcost}=19$ and ${\it newlocalcost}=24$. For $P_{2}$, ${\it localcost}=20$ and
	${\it newlocalcost}=20$. For $P_{3}$, ${\it localcost}=20$, whereas ${\it newlocalcost}=23$. So, only $P_{2}$ enters
	the IF statement that starts at line 32, and thus $P_{2}$ is the source processor. So, all the processors exit
	the WHILE loop that starts at line 17 (with ${\it COST}=83$).
\item[Update:] At this point, all the processors have ${\it COST}=83$. All the other variables remain unchanged, except for
	${\it sa}$, which becomes $3$.
\end{description}
\item[Step 4:] At the beginning of this step, ${\it OK}=0$ and ${\it nrt}=0$. Also, ${\it sa}=3$, ${\it myub}=0$, and
	${\it average}=3.45$. ${\it COST}$ and ${\it bestCOST}$ are both $83$.
\begin{description}
\item[Substep 1:] Suppose that $P_{1}$ selects $(1,1)$ and $(2,2)$, $P_{2}$ selects $(2,3)$ and $(4,4)$, and $P_{3}$ selects
	$(1,4)$ and $(4,4)$. For $P_{1}$, ${\it localcost}=20$ and ${\it newlocalcost}=22$. For $P_{2}$, ${\it localcost}=21$ and
	${\it newlocalcost}=24$. For $P_{3}$, ${\it localcost}=12$, whereas ${\it newlocalcost}=14$. So, none of the processors enters
	the IF statement that starts at line 32. Since ${\it nrt}<{\it MaxTrials1}$ at all the processors, none of the processors swaps the
	probes at its selected locations. So, all the processors remain with ${\it OK}=0$ at the end of this substep.
\item[Substep 2:] Suppose that $P_{1}$ selects $(1,4)$ and $(2,4)$, $P_{2}$ selects $(1,4)$ and $(2,2)$, and $P_{3}$ selects
	$(4,1)$ and $(2,1)$. For $P_{1}$, ${\it localcost}=16$ and ${\it newlocalcost}=17$. For $P_{2}$, ${\it localcost}=18$ and
	${\it newlocalcost}=22$. For $P_{3}$, ${\it localcost}=18$, whereas ${\it newlocalcost}=19$. So, none of the processors enters
	the IF statement that starts at line 32. Since ${\it nrt}={\it MaxTrials1}$ and the other conditions (at lines 42-43) are satisfied
	at all the processors, each of the processors swaps the probes at its selected locations.
	Suppose that $P_{1}$ is chosen to be the source processor at the synchronization that starts at line 47. Thus, at the end of this
	substep, all the processors have ${\it COST}=84$, and, since ${\it OK}=1$ at all the processors, all of them exit the WHILE loop
	that starts at line 17.
\item[Update:] At this point, each processors has ${\it COST}=84$. The ${\it average}$ variable becomes $3.50$, ${\it bestCOST}$
	and ${\it bestmicroarray}$ remain unchanged (since ${\it COST}$ just increased), ${\it sa}$ becomes $4$, and ${\it myub}$ remains $0$.
\end{description}
\item[Step 5:] Suppose that at this point, the time limit $T$ is exceeded at all the processors. Thus, all of them
	exit the WHILE loop that starts at line 14. Only processor $P_{1}$ returns the best cost (and the corresponding {\it bestmicroarray}
	configuration) found during the algorithm, which is $83$.
\end{description}
\end{example}
\subsection{ALG2}
	In this section we propose a variant of ALG1, called ALG2, which is given in Fig. \ref{fig:ALG2}. The only difference from ALG1 is that
	we have a new input parameter, namely ${\it MaxCost}$, which replaces the ${\it average}$ variable used in ALG1. This will allow ALG2 to
	give better results than ALG1 for some microarray dimensions.
\subsection{Results (on Randomly Generated Sets of Probes)}
	We have implemented the previous heuristics (TSP+1-Threading, epitaxial, row-epitaxial, recursive partitioning) and the algorithms
	discussed in this paper (LS,
	LS-Par, ALG1, and ALG2) on a SGI Altix machine with $64$ processors, using MPI \cite{mpi1}. Since the previous four heuristics
	(TSP+1-Threading, epitaxial,
	row-epitaxial, recursive partitioning) and the LS algorithm are sequential algorithms, we have run them using only one processor out
	of $64$ available. For LS-Par, ALG1, and ALG2, we have used all $64$ processors available in order to help them to converge faster
	to an approximate solution.
	For small microarrays, each of the four previous heuristics (TSP+1-Threading, epitaxial, row-epitaxial, recursive partitioning)
	takes just a few seconds. We can definitely use up to $64$ processors in order to reduce the time taken by each of them even further,
	but this does not help in reducing the cost. Indeed, the four previous heuristics (TSP+1-Threading, epitaxial, row-epitaxial,
	recursive partitioning), unlike LS, LS-Par, ALG1, and ALG2, are algorithms with a finite number of steps.
	For small microarrays of size at most $34\times{34}$, the epitaxial algorithm gives better results than TSP+1-Threading, row-epitaxial,
	and recursive partitioning. Thus, we compare the epitaxial algorithm against LS, LS-Par, ALG1, and ALG2.
	
	We have run LS with different probabilities, and collected results after $2$, $4$, $6$, $8$, and $10$ minutes. For LS-Par, we have
	used all $64$ processors available, and collected results after $2$, $4$, $6$, $8$, and $10$ minutes. For all microarray dimensions
	considered, LS and LS-Par perform worse than the epitaxial algorithm.
	We have also implemented ALG1 and ALG2 using all $64$ processors available, and collected results after $2$, $4$, $6$, $8$, and $10$
	minutes. The parameters used in order to get to the results shown in Tables \ref{table:ALG1results} and \ref{table:ALG2results}
	are as follows:
	${\it probelength}=25$,
	${\it MaxTrials1}=20$,
	${\it MaxTrials2}=40000$,
	${\it MaxCost}=160$,
	${\it MaxCost1}=10$,
	${\it MaxCost2}=10$,
	${\it winlength1}=1120$,
	${\it winlength2}=320$.
	For microarrays of size at most $32\times{32}$, ALG1 gives better results than the epitaxial algorithm. For microarrays
	of size $33\times{33}$ or more, ALG1 gives worse results than the epitaxial algorithm.
	For microarrays of size at most $34\times{34}$, ALG2 gives better results than the epitaxial algorithm. For microarrays
	of size $35\times{35}$ or more, ALG2 gives worse results than the epitaxial algorithm. We can also remark that ALG2
	performs better than ALG1. This suggests that the ${\it MaxCost}$ input parameter in ALG2 is more helpful than the ${\it average}$ variable
	in ALG1.


\begin{onecolumn}
\begin{figure}[ht]
\begin{center}
{\scriptsize
	\fbox{
		\begin{minipage}{415pt}
		\begin{tabbing}
		\hspace*{5mm}\=\hspace{5mm}\=\hspace{5mm}\=\hspace{5mm}\=\hspace{5mm}\=\hspace{5mm}\=  \kill
		{\tt Input:} ${\it SP}$, $T$, ${\it dim}$, ${\it probelength}$, and an input probability ${\it pr}$\\
		{\tt Output:} a microarray configuration as close as possible to the optimal\\
		\rule[3pt]{1.0\textwidth}{0.3pt}\\
		{\tt { 1: }}$\bullet$ Place the probes in ${\it SP}$ randomly on the {\it microarray}.\\
		{\tt { 2: }}$\bullet$ {\tt FOR} all $i\in\{1,\ldots,dim\}$ and all $j\in\{1,\ldots,dim\}$ {\tt DO}\\
		{\tt { 3: }}\hspace{27pt}$\bullet$ ${\it bestmicroarray}[i,j]\leftarrow{\it microarray}[i,j]$;\\
		{\tt { 4: }}$\bullet$ ${\it COST}\leftarrow{0}$;\\
		{\tt { 5: }}$\bullet$ {\tt FOR} all $i\in\{1,\ldots,dim\}$ and all $j\in\{1,\ldots,dim-1\}$ {\tt DO}\\
		{\tt { 6: }}\hspace{27pt}$\bullet$ ${\it COST}\leftarrow{\it COST}+{\it HammingDistance}({\it microarray}[i,j],{\it microarray}[i,j+1])$;\\
		{\tt { 7: }}$\bullet$ {\tt FOR} all $j\in\{1,\ldots,dim\}$ and all $i\in\{1,\ldots,dim-1\}$ {\tt DO}\\
		{\tt { 8: }}\hspace{27pt}$\bullet$ ${\it COST}\leftarrow{\it COST}+{\it HammingDistance}({\it microarray}[i,j],{\it microarray}[i+1,j])$;\\
		{\tt { 9: }}$\bullet$ ${\it bestCOST}\leftarrow{\it COST}$;\\
		{\tt {10: }}$\bullet$ {\tt WHILE} (the time taken by the algorithm is $\leq$ $T$) {\tt DO}\\
		{\tt {11: }}\hspace{27pt}$\bullet$ Choose two random locations on the {\it microarray}, say $(l_{1},c_{1})$ and $(l_{2},c_{2})$. (The random locations that\\
		{\tt {12: }}\hspace{34pt}are chosen depend on the current processor, since each of the processors has its own random number\\
		{\tt {13: }}\hspace{34pt}generator.)\\
		{\tt {14: }}\hspace{27pt}$\bullet$ Let ${\it localcost1}$ be the sum of the Hamming distances between the probe ${\it microarray}[l_{1},c_{1}]$ and the\\
		{\tt {15: }}\hspace{34pt}probes that are currently neighbors to $(l_{1},c_{1})$.\\
		{\tt {16: }}\hspace{27pt}$\bullet$ Let ${\it localcost2}$ be the sum of the Hamming distances between the probe ${\it microarray}[l_{2},c_{2}]$ and the\\
		{\tt {17: }}\hspace{34pt}probes that are currently neighbors to $(l_{2},c_{2})$.\\
		{\tt {18: }}\hspace{27pt}$\bullet$ Let ${\it newlocalcost1}$ be the sum of the Hamming distances between the probe ${\it microarray}[l_{2},c_{2}]$ and\\
		{\tt {19: }}\hspace{34pt}the probes that are currently neighbors to $(l_{1},c_{1})$.\\
		{\tt {20: }}\hspace{27pt}$\bullet$ Let ${\it newlocalcost2}$ be the sum of the Hamming distances between the probe ${\it microarray}[l_{1},c_{1}]$ and\\
		{\tt {21: }}\hspace{34pt}the probes that are currently neighbors to $(l_{2},c_{2})$.\\
		{\tt {22: }}\hspace{27pt}$\bullet$ ${\it localcost}\leftarrow{\it localcost1}+{\it localcost2}$;\\
		{\tt {23: }}\hspace{27pt}$\bullet$ ${\it newlocalcost}\leftarrow{\it newlocalcost1}+{\it newlocalcost2}$;\\
		{\tt {24: }}\hspace{27pt}$\bullet$ {\tt IF} (${\it newlocalcost}<{\it localcost}$) {\tt THEN}\\
		{\tt {25: }}\hspace{54pt}$\bullet$ Swap the probes at locations $(l_{1},c_{1})$ and $(l_{2},c_{2})$;\\
		{\tt {26: }}\hspace{54pt}$\bullet$ ${\it COST}\leftarrow{\it COST}-({\it localcost}-{\it newlocalcost})$;\\
		{\tt {27: }}\hspace{54pt}$\bullet$ ${\it bestCOST}\leftarrow{\it COST}$;\\
		{\tt {28: }}\hspace{54pt}$\bullet$ {\tt FOR} all $i\in\{1,\ldots,dim\}$ and all $j\in\{1,\ldots,dim\}$ {\tt DO}\\
		{\tt {29: }}\hspace{81pt}$\bullet$ ${\it bestmicroarray}[i,j]\leftarrow{\it microarray}[i,j]$;\\
		{\tt {30: }}\hspace{34pt}{\tt ELSE}\\
		{\tt {31: }}\hspace{54pt}$\bullet$ Randomly generate a number $N$ in the interval $[0,1]$.\\
		{\tt {32: }}\hspace{54pt}$\bullet$ {\tt IF} $N\leq{\it pr}$ {\tt THEN}\\
		{\tt {33: }}\hspace{81pt}$\bullet$ Swap the probes at locations $(l_{1},c_{1})$ and $(l_{2},c_{2})$;\\
		{\tt {34: }}\hspace{81pt}$\bullet$ ${\it COST}\leftarrow{\it COST}+({\it newlocalcost}-{\it localcost})$;\\
		{\tt {35: }}$\bullet$ return ${\it bestmicroarray}$;
		\end{tabbing}
		\end{minipage}
	}
}
\end{center}
\caption{LS: a local-search-based sequential algorithm for the BLMP}
\label{fig:LS}
\end{figure}
\begin{figure}[ht]
\begin{center}
{\scriptsize
	\fbox{
		\begin{minipage}{415pt}
		\begin{tabbing}
		\hspace*{5mm}\=\hspace{5mm}\=\hspace{5mm}\=\hspace{5mm}\=\hspace{5mm}\=\hspace{5mm}\=  \kill
		{\tt Input:} ${\it SP}$, $T$, ${\it dim}$, ${\it probelength}$, and an input probability ${\it pr}$\\
		{\tt Output:} a microarray configuration as close as possible to the optimal\\
		\rule[3pt]{1.0\textwidth}{0.3pt}\\
		{\tt { 1: }}$\bullet$ Processor $P_{1}$ places the probes in ${\it SP}$ randomly on its {\it microarray} and then sends its microarray configuration\\
		{\tt { 2: }}\hspace{7pt}to all the other processors.\\
		{\tt { 3: }}$\bullet$ {\tt FOR} all $i\in\{1,\ldots,dim\}$ and all $j\in\{1,\ldots,dim\}$ {\tt DO}\\
		{\tt { 4: }}\hspace{27pt}$\bullet$ ${\it bestmicroarray}[i,j]\leftarrow{\it microarray}[i,j]$;\\
		{\tt { 5: }}$\bullet$ ${\it COST}\leftarrow{0}$;\\
		{\tt { 6: }}$\bullet$ {\tt FOR} all $i\in\{1,\ldots,dim\}$ and all $j\in\{1,\ldots,dim-1\}$ {\tt DO}\\
		{\tt { 7: }}\hspace{27pt}$\bullet$ ${\it COST}\leftarrow{\it COST}+{\it HammingDistance}({\it microarray}[i,j],{\it microarray}[i,j+1])$;\\
		{\tt { 8: }}$\bullet$ {\tt FOR} all $j\in\{1,\ldots,dim\}$ and all $i\in\{1,\ldots,dim-1\}$ {\tt DO}\\
		{\tt { 9: }}\hspace{27pt}$\bullet$ ${\it COST}\leftarrow{\it COST}+{\it HammingDistance}({\it microarray}[i,j],{\it microarray}[i+1,j])$;\\
		{\tt {10: }}$\bullet$ ${\it bestCOST}\leftarrow{\it COST}$;\\
		{\tt {11: }}$\bullet$ {\tt WHILE} (the time taken by the algorithm is $\leq$ $T$) {\tt DO}\\
		{\tt {12: }}\hspace{27pt}$\bullet$ Choose two random locations on the {\it microarray}, say $(l_{1},c_{1})$ and $(l_{2},c_{2})$. (The random locations that\\
		{\tt {13: }}\hspace{34pt}are chosen depend on the current processor, since each of the processors has its own random number\\
		{\tt {14: }}\hspace{34pt}generator.)\\
		{\tt {15: }}\hspace{27pt}$\bullet$ Let ${\it localcost1}$ be the sum of the Hamming distances between the probe ${\it microarray}[l_{1},c_{1}]$ and the\\
		{\tt {16: }}\hspace{34pt}probes that are currently neighbors to $(l_{1},c_{1})$.\\
		{\tt {17: }}\hspace{27pt}$\bullet$ Let ${\it localcost2}$ be the sum of the Hamming distances between the probe ${\it microarray}[l_{2},c_{2}]$ and the\\
		{\tt {18: }}\hspace{34pt}probes that are currently neighbors to $(l_{2},c_{2})$.\\
		{\tt {19: }}\hspace{27pt}$\bullet$ Let ${\it newlocalcost1}$ be the sum of the Hamming distances between the probe ${\it microarray}[l_{2},c_{2}]$ and\\
		{\tt {20: }}\hspace{34pt}the probes that are currently neighbors to $(l_{1},c_{1})$.\\
		{\tt {21: }}\hspace{27pt}$\bullet$ Let ${\it newlocalcost2}$ be the sum of the Hamming distances between the probe ${\it microarray}[l_{1},c_{1}]$ and\\
		{\tt {22: }}\hspace{34pt}the probes that are currently neighbors to $(l_{2},c_{2})$.\\
		{\tt {23: }}\hspace{27pt}$\bullet$ ${\it localcost}\leftarrow{\it localcost1}+{\it localcost2}$;\\
		{\tt {24: }}\hspace{27pt}$\bullet$ ${\it newlocalcost}\leftarrow{\it newlocalcost1}+{\it newlocalcost2}$;\\
		{\tt {25: }}\hspace{27pt}$\bullet$ {\tt IF} (${\it newlocalcost}<{\it localcost}$) {\tt THEN}\\
		{\tt {26: }}\hspace{54pt}$\bullet$ Swap the probes at locations $(l_{1},c_{1})$ and $(l_{2},c_{2})$;\\
		{\tt {27: }}\hspace{54pt}$\bullet$ ${\it COST}\leftarrow{\it COST}-({\it localcost}-{\it newlocalcost})$;\\
		{\tt {28: }}\hspace{54pt}$\bullet$ ${\it bestCOST}\leftarrow{\it COST}$;\\
		{\tt {29: }}\hspace{54pt}$\bullet$ {\tt FOR} all $i\in\{1,\ldots,dim\}$ and all $j\in\{1,\ldots,dim\}$ {\tt DO}\\
		{\tt {30: }}\hspace{81pt}$\bullet$ ${\it bestmicroarray}[i,j]\leftarrow{\it microarray}[i,j]$;\\
		{\tt {31: }}\hspace{34pt}{\tt ELSE}\\
		{\tt {32: }}\hspace{54pt}$\bullet$ Randomly generate a number $N$ in the interval $[0,1]$.\\
		{\tt {33: }}\hspace{54pt}$\bullet$ {\tt IF} $N\leq{\it pr}$ {\tt THEN}\\
		{\tt {34: }}\hspace{81pt}$\bullet$ Swap the probes at locations $(l_{1},c_{1})$ and $(l_{2},c_{2})$;\\
		{\tt {35: }}\hspace{81pt}$\bullet$ ${\it COST}\leftarrow{\it COST}+({\it newlocalcost}-{\it localcost})$;\\
		{\tt {36: }}\hspace{27pt}$\bullet$ Processors synchronize with each other. Let $P_{\it source}$ be randomly selected out of those processors that\\
		{\tt {37: }}\hspace{34pt}have a minimum ${\it COST}$. All the other processors update their ${\it microarray}$, ${\it COST}$, ${\it bestmicroarray}$,\\
		{\tt {38: }}\hspace{34pt}${\it bestCOST}$ with the corresponding variables from $P_{\it source}$.\\
		{\tt {39: }}$\bullet$ Processor $P_{1}$ returns ${\it bestmicroarray}$;
		\end{tabbing}
		\end{minipage}
	}
}
\end{center}
\caption{LS-Par: a local-search-based parallel algorithm for the BLMP (The pseudocode shown here is executed by each of the processors; only
processor $P_{1}$ returns the final result.)}
\label{fig:LS-Par}
\end{figure}
\begin{figure}[ht]
\setlength{\unitlength}{1pt}
{\scriptsize
\begin{picture}(468,84)(0,0)
	\put(193,82){\line(1,0){80}}
	\put(273,82){\line(0,-1){80}}
	\put(273,2){\line(-1,0){80}}
	\put(193,2){\line(0,1){80}}
	
	\put(193,22){\line(1,0){80}}
	\put(193,42){\line(1,0){80}}
	\put(193,62){\line(1,0){80}}
	
	\put(213,2){\line(0,1){80}}
	\put(233,2){\line(0,1){80}}
	\put(253,2){\line(0,1){80}}
	
	\put(199,70){$p_{1}$}
	\put(219,70){$p_{2}$}
	\put(239,70){$p_{3}$}
	\put(259,70){$p_{4}$}
	
	\put(199,50){$p_{5}$}
	\put(219,50){$p_{6}$}
	\put(239,50){$p_{7}$}
	\put(259,50){$p_{8}$}
	
	\put(199,30){$p_{9}$}
	\put(219,30){$p_{10}$}
	\put(239,30){$p_{11}$}
	\put(259,30){$p_{12}$}
	
	\put(199,10){$p_{13}$}
	\put(219,10){$p_{14}$}
	\put(239,10){$p_{15}$}
	\put(259,10){$p_{16}$}
\end{picture}
}
\caption{The initial placement (at all the processors) of the probes on the microarray}
\label{fig:InitialPlacement}
\end{figure}
\begin{figure}
\begin{center}
{\scriptsize
	\fbox{
		\begin{minipage}{415pt}
		\begin{tabbing}
		\hspace*{5mm}\=\hspace{5mm}\=\hspace{5mm}\=\hspace{5mm}\=\hspace{5mm}\=\hspace{5mm}\=  \kill
		{\tt Input:} ${\it SP}$, $T$, ${\it dim}$, ${\it probelength}$, ${\it MaxTrials1}$, ${\it MaxTrials2}$, ${\it MaxCost1}$, ${\it MaxCost2}$, ${\it winlength1}$, ${\it winlength2}$\\
		{\tt Output:} a microarray configuration as close as possible to the optimal\\
		\rule[3pt]{1.0\textwidth}{0.3pt}\\
		{\tt { 1: }}$\bullet$ Processor $P_{1}$ places all the probes in ${\it SP}$ randomly on its {\it microarray}, and then sends its {\it microarray} to all\\
		{\tt { 2: }}\hspace{7pt}the other processors;\\
		{\tt { 3: }}$\bullet$ {\tt FOR} all $i\in\{1,\ldots,dim\}$ and all $j\in\{1,\ldots,dim\}$ {\tt DO}\\
		{\tt { 4: }}\hspace{27pt}$\bullet$ ${\it bestmicroarray}[i,j]\leftarrow{\it microarray}[i,j]$;\\
		{\tt { 5: }}$\bullet$ ${\it COST}\leftarrow{0}$;\\
		{\tt { 6: }}$\bullet$ {\tt FOR} all $i\in\{1,\ldots,dim\}$ and all $j\in\{1,\ldots,dim-1\}$ {\tt DO}\\
		{\tt { 7: }}\hspace{27pt}$\bullet$ ${\it COST}\leftarrow{\it COST}+{\it HammingDistance}({\it microarray}[i,j],{\it microarray}[i,j+1])$;\\
		{\tt { 8: }}$\bullet$ {\tt FOR} all $j\in\{1,\ldots,dim\}$ and all $i\in\{1,\ldots,dim-1\}$ {\tt DO}\\
		{\tt { 9: }}\hspace{27pt}$\bullet$ ${\it COST}\leftarrow{\it COST}+{\it HammingDistance}({\it microarray}[i,j],{\it microarray}[i+1,j])$;\\
		{\tt {10: }}$\bullet$ ${\it average}\leftarrow{[{\it COST}/({\it dim}*({\it dim}-1)*2)]}$;\\
		{\tt {11: }}$\bullet$ ${\it bestCOST}\leftarrow{\it COST}$;\\
		{\tt {12: }}$\bullet$ ${\it sa}\leftarrow{0}$;\\
		{\tt {13: }}$\bullet$ ${\it myub}\leftarrow{0}$;\\
		{\tt {14: }}$\bullet$ {\tt WHILE} (the time taken by the algorithm is $\leq$ $T$) {\tt DO}\\
		{\tt {15: }}\hspace{27pt}$\bullet$ ${\it OK}\leftarrow{0};$\\
		{\tt {16: }}\hspace{27pt}$\bullet$ ${\it nrt}\leftarrow{0};$\\
		{\tt {17: }}\hspace{27pt}$\bullet$ {\tt WHILE} (${\it OK}=0$) {\tt DO}\\
		{\tt {18: }}\hspace{54pt}$\bullet$ ${\it nrt}\leftarrow{\it nrt}+1;$\\
		{\tt {19: }}\hspace{54pt}$\bullet$ Choose two random locations on the {\it microarray}, say $(l_{1},c_{1})$ and $(l_{2},c_{2})$. (The random\\
		{\tt {20: }}\hspace{61pt}locations that are chosen depend on the current processor, since each processor has its\\
		{\tt {21: }}\hspace{61pt}own random number generator.)\\
		{\tt {22: }}\hspace{54pt}$\bullet$ Let ${\it localcost1}$ be the sum of the Hamming distances between the probe ${\it microarray}[l_{1},c_{1}]$\\
		{\tt {23: }}\hspace{61pt}and the probes that are currently neighbors to $(l_{1},c_{1})$.\\
		{\tt {24: }}\hspace{54pt}$\bullet$ Let ${\it localcost2}$ be the sum of the Hamming distances between the probe ${\it microarray}[l_{2},c_{2}]$\\
		{\tt {25: }}\hspace{61pt}and the probes that are currently neighbors to $(l_{2},c_{2})$.\\
		{\tt {26: }}\hspace{54pt}$\bullet$ Let ${\it newlocalcost1}$ be the sum of the Hamming distances between the probe ${\it microarray}[l_{2},c_{2}]$\\
		{\tt {27: }}\hspace{61pt}and the probes that are currently neighbors to $(l_{1},c_{1})$.\\
		{\tt {28: }}\hspace{54pt}$\bullet$ Let ${\it newlocalcost2}$ be the sum of the Hamming distances between the probe ${\it microarray}[l_{1},c_{1}]$\\
		{\tt {29: }}\hspace{61pt}that are currently neighbors to $(l_{2},c_{2})$.\\
		{\tt {30: }}\hspace{54pt}$\bullet$ ${\it localcost}\leftarrow{\it localcost1}+{\it localcost2}$;\\
		{\tt {31: }}\hspace{54pt}$\bullet$ ${\it newlocalcost}\leftarrow{\it newlocalcost1}+{\it newlocalcost2}$;\\
		{\tt {32: }}\hspace{54pt}$\bullet$ {\tt IF} (${\it newlocalcost}\leq{\it localcost}$) {\tt THEN}\\
		{\tt {33: }}\hspace{81pt}$\bullet$ ${\it OK}\leftarrow{1};$\\
		{\tt {34: }}\hspace{81pt}$\bullet$ Swap the probes at locations $(l_{1},c_{1})$ and $(l_{2},c_{2})$;\\
		{\tt {35: }}\hspace{81pt}$\bullet$ ${\it COST}\leftarrow{\it COST}-({\it localcost}-{\it newlocalcost})$;\\
		{\tt {36: }}\hspace{54pt}$\bullet$ Processors synchronize with each other. If at least one of them has ${\it OK}=1$, then let $P_{\it source}$\\
		{\tt {37: }}\hspace{61pt}be one of those processors with ${\it OK}=1$, randomly chosen. All the other processors update\\
		{\tt {38: }}\hspace{61pt}their {\it microarray} and ${\it COST}$ with the corresponding variables from $P_{\it source}$, and then set their\\
		{\tt {39: }}\hspace{61pt}${\it OK}$ their variable to $1$.\\
		{\tt {40: }}\hspace{54pt}$\bullet$ {\tt IF} ($OK=0$) {\tt THEN}\\
		{\tt {41: }}\hspace{81pt}$\bullet$ {\tt IF} $({\it nrt}\geq{\it MaxTrials1})$ {\tt THEN}\\
		{\tt {42: }}\hspace{108pt}$\bullet$ {\tt IF} $({\it newlocalcost}\leq\lfloor{8*}{\it average}\rfloor+{\it myub})$ {\tt THEN}\\
		{\tt {43: }}\hspace{135pt}$\bullet$ {\tt IF} $({\it COST}+{\it newlocalcost}-{\it localcost}\leq{\it bestCOST}+{\it MaxCost2})$ {\tt THEN}\\
		{\tt {44: }}\hspace{162pt}$\bullet$ ${\it OK}\leftarrow{1};$\\
		{\tt {45: }}\hspace{162pt}$\bullet$ Swap the probes at locations $(l_{1},c_{1})$ and $(l_{2},c_{2})$;\\
		{\tt {46: }}\hspace{162pt}$\bullet$ ${\it COST}\leftarrow{\it COST}+({\it newlocalcost}-{\it localcost})$;\\
		{\tt {47: }}\hspace{81pt}$\bullet$ Processors synchronize with each other. If at least one of them has ${\it OK}=1$, then let\\
		{\tt {48: }}\hspace{88pt}$P_{\it source}$ be one of those processors with ${\it OK}=1$, randomly chosen. All the other\\
		{\tt {49: }}\hspace{88pt}processors update their {\it microarray} and ${\it COST}$ with the corresponding variables from\\
		{\tt {50: }}\hspace{88pt}$P_{\it source}$, and then set their ${\it OK}$ variable to $1$.\\
		{\tt {51: }}\hspace{54pt}$\bullet$ {\tt IF} $({\it OK}=0)$ and $({\it nrt}\geq{\it MaxTrials2})$ {\tt THEN}\\
		{\tt {52: }}\hspace{81pt}$\bullet$ {\tt BREAK} the {\tt WHILE} that starts at line 17;\\
		{\tt {53: }}\hspace{27pt}$\bullet$  ${\it average}\leftarrow{[{\it COST}/({\it dim}*({\it dim}-1)*2)]}$;\\
		{\tt {54: }}\hspace{27pt}$\bullet$ {\tt IF} $({\it COST}<{\it bestCOST})$ {\tt THEN}\\
		{\tt {55: }}\hspace{54pt}$\bullet$ ${\it bestCOST}\leftarrow{\it COST}$;\\
		{\tt {56: }}\hspace{54pt}$\bullet$ {\tt FOR} all $i\in\{1,\ldots,dim\}$ and all $j\in\{1,\ldots,dim\}$ {\tt DO}\\
		{\tt {57: }}\hspace{81pt}$\bullet$ ${\it bestmicroarray}[i,j]\leftarrow{\it microarray}[i,j]$;\\
		{\tt {58: }}\hspace{27pt}$\bullet$ ${\it sa}\leftarrow{\it sa}+1$;\\
		{\tt {59: }}\hspace{27pt}$\bullet$ {\tt IF} $({\it myub}=0)$ {\tt THEN}\\
		{\tt {60: }}\hspace{54pt}$\bullet$ {\tt IF} $({\it sa}={\it winlength1})$ {\tt THEN}\\
		{\tt {61: }}\hspace{81pt}$\bullet$ ${\it sa}\leftarrow{0}$;\\
		{\tt {62: }}\hspace{81pt}$\bullet$ ${\it myub}\leftarrow{\it MaxCost1}$;\\
		{\tt {63: }}\hspace{34pt}{\tt ELSE}\\
		{\tt {64: }}\hspace{54pt}$\bullet$ {\tt IF} $({\it sa}={\it winlength}2)$ {\tt THEN}\\
		{\tt {65: }}\hspace{81pt}$\bullet$ ${\it sa}\leftarrow{0}$;\\
		{\tt {66: }}\hspace{81pt}$\bullet$ ${\it myub}\leftarrow{0}$;\\
		{\tt {67: }}$\bullet$ Processor $P_{1}$ returns ${\it bestmicroarray}$;
		\end{tabbing}
		\end{minipage}
	}
}
\end{center}
\caption{ALG1 (the pseudocode shown here is executed by each of the processors involved in the algorithm; only processor $P_{1}$ returns the final
result.)}
\label{fig:ALG1}
\end{figure}
\begin{figure}
\begin{center}
{\scriptsize
	\fbox{
		\begin{minipage}{415pt}
		\begin{tabbing}
		\hspace*{5mm}\=\hspace{5mm}\=\hspace{5mm}\=\hspace{5mm}\=\hspace{5mm}\=\hspace{5mm}\=  \kill
		{\tt Input:} ${\it SP}$, $T$, ${\it dim}$, ${\it probelength}$, ${\it MaxTrials1}$, ${\it MaxTrials2}$, ${\it MaxCost}$, ${\it MaxCost1}$, ${\it MaxCost2}$, ${\it winlength1}$, ${\it winlength2}$\\
		{\tt Output:} a microarray configuration as close as possible to the optimal\\
		\rule[3pt]{1.0\textwidth}{0.3pt}\\
		{\tt { 1: }}$\bullet$ Processor $P_{1}$ places all the probes in ${\it SP}$ randomly on its {\it microarray}, and then sends its {\it microarray} to all\\
		{\tt { 2: }}\hspace{7pt}the other processors;\\
		{\tt { 3: }}$\bullet$ {\tt FOR} all $i\in\{1,\ldots,dim\}$ and all $j\in\{1,\ldots,dim\}$ {\tt DO}\\
		{\tt { 4: }}\hspace{27pt}$\bullet$ ${\it bestmicroarray}[i,j]\leftarrow{\it microarray}[i,j]$;\\
		{\tt { 5: }}$\bullet$ ${\it COST}\leftarrow{0}$;\\
		{\tt { 6: }}$\bullet$ {\tt FOR} all $i\in\{1,\ldots,dim\}$ and all $j\in\{1,\ldots,dim-1\}$ {\tt DO}\\
		{\tt { 7: }}\hspace{27pt}$\bullet$ ${\it COST}\leftarrow{\it COST}+{\it HammingDistance}({\it microarray}[i,j],{\it microarray}[i,j+1])$;\\
		{\tt { 8: }}$\bullet$ {\tt FOR} all $j\in\{1,\ldots,dim\}$ and all $i\in\{1,\ldots,dim-1\}$ {\tt DO}\\
		{\tt { 9: }}\hspace{27pt}$\bullet$ ${\it COST}\leftarrow{\it COST}+{\it HammingDistance}({\it microarray}[i,j],{\it microarray}[i+1,j])$;\\
		{\tt {10: }}$\bullet$ ${\it bestCOST}\leftarrow{\it COST}$;\\
		{\tt {11: }}$\bullet$ ${\it sa}\leftarrow{0}$;\\
		{\tt {12: }}$\bullet$ ${\it myub}\leftarrow{0}$;\\
		{\tt {13: }}$\bullet$ {\tt WHILE} (the time taken by the algorithm is $\leq$ $T$) {\tt DO}\\
		{\tt {14: }}\hspace{27pt}$\bullet$ ${\it OK}\leftarrow{0};$\\
		{\tt {15: }}\hspace{27pt}$\bullet$ ${\it nrt}\leftarrow{0};$\\
		{\tt {16: }}\hspace{27pt}$\bullet$ {\tt WHILE} (${\it OK}=0$) {\tt DO}\\
		{\tt {17: }}\hspace{54pt}$\bullet$ ${\it nrt}\leftarrow{\it nrt}+1;$\\
		{\tt {18: }}\hspace{54pt}$\bullet$ Choose two random locations on the {\it microarray}, say $(l_{1},c_{1})$ and $(l_{2},c_{2})$. (The random\\
		{\tt {19: }}\hspace{61pt}locations that are chosen depend on the current processor, since each processor has its\\
		{\tt {20: }}\hspace{61pt}own random number generator.)\\
		{\tt {21: }}\hspace{54pt}$\bullet$ Let ${\it localcost1}$ be the sum of the Hamming distances between the probe ${\it microarray}[l_{1},c_{1}]$\\
		{\tt {22: }}\hspace{61pt}and the probes that are currently neighbors to $(l_{1},c_{1})$.\\
		{\tt {23: }}\hspace{54pt}$\bullet$ Let ${\it localcost2}$ be the sum of the Hamming distances between the probe ${\it microarray}[l_{2},c_{2}]$\\
		{\tt {24: }}\hspace{61pt}and the probes that are currently neighbors to $(l_{2},c_{2})$.\\
		{\tt {25: }}\hspace{54pt}$\bullet$ Let ${\it newlocalcost1}$ be the sum of the Hamming distances between the probe ${\it microarray}[l_{2},c_{2}]$\\
		{\tt {26: }}\hspace{61pt}and the probes that are currently neighbors to $(l_{1},c_{1})$.\\
		{\tt {27: }}\hspace{54pt}$\bullet$ Let ${\it newlocalcost2}$ be the sum of the Hamming distances between the probe ${\it microarray}[l_{1},c_{1}]$\\
		{\tt {28: }}\hspace{61pt}that are currently neighbors to $(l_{2},c_{2})$.\\
		{\tt {29: }}\hspace{54pt}$\bullet$ ${\it localcost}\leftarrow{\it localcost1}+{\it localcost2}$;\\
		{\tt {30: }}\hspace{54pt}$\bullet$ ${\it newlocalcost}\leftarrow{\it newlocalcost1}+{\it newlocalcost2}$;\\
		{\tt {31: }}\hspace{54pt}$\bullet$ {\tt IF} (${\it newlocalcost}\leq{\it localcost}$) {\tt THEN}\\
		{\tt {32: }}\hspace{81pt}$\bullet$ ${\it OK}\leftarrow{1};$\\
		{\tt {33: }}\hspace{81pt}$\bullet$ Swap the probes at locations $(l_{1},c_{1})$ and $(l_{2},c_{2})$;\\
		{\tt {34: }}\hspace{81pt}$\bullet$ ${\it COST}\leftarrow{\it COST}-({\it localcost}-{\it newlocalcost})$;\\
		{\tt {35: }}\hspace{54pt}$\bullet$ Processors synchronize with each other. If at least one of them has ${\it OK}=1$, then let $P_{\it source}$\\
		{\tt {36: }}\hspace{61pt}be one of those processors with ${\it OK}=1$, randomly chosen. All the other processors update\\
		{\tt {37: }}\hspace{61pt}their {\it microarray} and ${\it COST}$ with the corresponding variables from $P_{\it source}$, and then set their\\
		{\tt {38: }}\hspace{61pt}${\it OK}$ their variable to $1$.\\
		{\tt {39: }}\hspace{54pt}$\bullet$ {\tt IF} ($OK=0$) {\tt THEN}\\
		{\tt {40: }}\hspace{81pt}$\bullet$ {\tt IF} $({\it nrt}\geq{\it MaxTrials1})$ {\tt THEN}\\
		{\tt {41: }}\hspace{108pt}$\bullet$ {\tt IF} $({\it newlocalcost}\leq{\it MaxCost}+{\it myub})$ {\tt THEN}\\
		{\tt {42: }}\hspace{135pt}$\bullet$ {\tt IF} $({\it COST}+{\it newlocalcost}-{\it localcost}\leq{\it bestCOST}+{\it MaxCost2})$ {\tt THEN}\\
		{\tt {43: }}\hspace{162pt}$\bullet$ ${\it OK}\leftarrow{1};$\\
		{\tt {44: }}\hspace{162pt}$\bullet$ Swap the probes at locations $(l_{1},c_{1})$ and $(l_{2},c_{2})$;\\
		{\tt {45: }}\hspace{162pt}$\bullet$ ${\it COST}\leftarrow{\it COST}+({\it newlocalcost}-{\it localcost})$;\\
		{\tt {46: }}\hspace{81pt}$\bullet$ Processors synchronize with each other. If at least one of them has ${\it OK}=1$, then let\\
		{\tt {47: }}\hspace{88pt}$P_{\it source}$ be one of those processors with ${\it OK}=1$, randomly chosen. All the other\\
		{\tt {48: }}\hspace{88pt}processors update their {\it microarray} and ${\it COST}$ with the corresponding variables from\\
		{\tt {49: }}\hspace{88pt}$P_{\it source}$, and then set their ${\it OK}$ variable to $1$.\\
		{\tt {50: }}\hspace{54pt}$\bullet$ {\tt IF} $({\it OK}=0)$ and $({\it nrt}\geq{\it MaxTrials2})$ {\tt THEN}\\
		{\tt {51: }}\hspace{81pt}$\bullet$ {\tt BREAK} the {\tt WHILE} that starts at line 16;\\
		{\tt {52: }}\hspace{27pt}$\bullet$ {\tt IF} $({\it COST}<{\it bestCOST})$ {\tt THEN}\\
		{\tt {53: }}\hspace{54pt}$\bullet$ ${\it bestCOST}\leftarrow{\it COST}$;\\
		{\tt {54: }}\hspace{54pt}$\bullet$ {\tt FOR} all $i\in\{1,\ldots,dim\}$ and all $j\in\{1,\ldots,dim\}$ {\tt DO}\\
		{\tt {55: }}\hspace{81pt}$\bullet$ ${\it bestmicroarray}[i,j]\leftarrow{\it microarray}[i,j]$;\\
		{\tt {56: }}\hspace{27pt}$\bullet$ ${\it sa}\leftarrow{\it sa}+1$;\\
		{\tt {57: }}\hspace{27pt}$\bullet$ {\tt IF} $({\it myub}=0)$ {\tt THEN}\\
		{\tt {58: }}\hspace{54pt}$\bullet$ {\tt IF} $({\it sa}={\it winlength1})$ {\tt THEN}\\
		{\tt {59: }}\hspace{81pt}$\bullet$ ${\it sa}\leftarrow{0}$;\\
		{\tt {60: }}\hspace{81pt}$\bullet$ ${\it myub}\leftarrow{\it MaxCost1}$;\\
		{\tt {61: }}\hspace{34pt}{\tt ELSE}\\
		{\tt {62: }}\hspace{54pt}$\bullet$ {\tt IF} $({\it sa}={\it winlength}2)$ {\tt THEN}\\
		{\tt {63: }}\hspace{81pt}$\bullet$ ${\it sa}\leftarrow{0}$;\\
		{\tt {64: }}\hspace{81pt}$\bullet$ ${\it myub}\leftarrow{0}$;\\
		{\tt {65: }}$\bullet$ Processor $P_{1}$ returns ${\it bestmicroarray}$;
		\end{tabbing}
		\end{minipage}
	}
}
\end{center}
\caption{ALG2 (the pseudocode shown here is executed by each of the processors involved in the algorithm; only processor $P_{1}$ returns the final
result.)}
\label{fig:ALG2}
\end{figure}


\begin{table}
\caption{The $16$ probes in ${\it SP}$}
\centering
{\scriptsize
\label{table:Probes}
\begin{tabular}{|rr|}
\hline
$p_{1}= {\tt CGATT}$	&   \hspace{10pt}$p_{9}= {\tt ATACG}$			\\
$p_{2}= {\tt GGGCC}$	&	\hspace{10pt}$p_{10}={\tt CCCTC}$			\\
$p_{3}= {\tt ATCGA}$	&	\hspace{10pt}$p_{11}={\tt GGAGA}$			\\
$p_{4}= {\tt ATGTC}$	&	\hspace{10pt}$p_{12}={\tt AGCCG}$			\\
$p_{5}= {\tt TTAGT}$	&	\hspace{10pt}$p_{13}={\tt AGACA}$			\\
$p_{6}= {\tt ACCAG}$	&	\hspace{10pt}$p_{14}={\tt ACCTA}$			\\
$p_{7}= {\tt CCCGA}$	&	\hspace{10pt}$p_{15}={\tt GAATC}$			\\
$p_{8}= {\tt AATTC}$	&	\hspace{10pt}$p_{16}={\tt GATTT}$			\\\hline
\end{tabular}
}
\end{table}
\begin{table}
\caption{The Hamming distances between every two probes in ${\it SP}$}
\centering
{\scriptsize
\label{table:HammingDistances}
\begin{tabular}{|r|r|r|r|r|r|r|r|r|r|r|r|r|r|r|r|r|}
\hline
		& \hspace{2pt}$p_{1}$\hspace{2pt} & \hspace{2pt}$p_{2}$\hspace{2pt} & \hspace{2pt}$p_{3}$\hspace{2pt} & \hspace{2pt}$p_{4}$\hspace{2pt} & \hspace{2pt}$p_{5}$\hspace{2pt} & \hspace{2pt}$p_{6}$\hspace{2pt} & \hspace{2pt}$p_{7}$\hspace{2pt} & \hspace{2pt}$p_{8}$\hspace{2pt} & \hspace{2pt}$p_{9}$\hspace{2pt} & $p_{10}$ & $p_{11}$ & $p_{12}$ & $p_{13}$ & $p_{14}$ & $p_{15}$ & $p_{16}$\\\hline
$p_{1}$	& 		  &	4		& 5		  &	4		& 3		  & 5		& 4		  &	4		& 4		  & 3		 & 3		& 4		   & 3		  &	4		 & 3		& 3	      \\\hline
$p_{2}$	& 		  & 		& 5		  &	3		& 5		  &	5		& 5		  &	4		& 4		  &	4		 & 3		& 3		   & 3		  &	5		 & 3		& 4	      \\\hline
$p_{3}$	& 		  & 		&		  &	3		& 3		  &	3		& 2		  &	4		& 3		  &	4		 & 3		& 3		   & 3		  &	2		 & 5		& 5	      \\\hline
$p_{4}$	& 		  &			&		  &			& 4		  & 4		& 5		  &	2		& 3		  &	3		 & 5		& 4		   & 4		  &	3		 & 3		& 4	      \\\hline
$p_{5}$	& 		  & 		&		  &			&		  &	5		& 4		  &	5		& 3		  &	5		 & 3		& 5		   & 4		  &	5		 & 4		& 4	      \\\hline
$p_{6}$	& 		  & 		&		  &			&		  &			& 3		  &	4		& 3		  &	3		 & 5		& 2		   & 4		  &	2		 & 5		& 5	      \\\hline
$p_{7}$	& 		  &			&		  &			&		  &			&		  &	5		& 5		  &	2		 & 3		& 4		   & 4		  &	2		 & 5		& 5	      \\\hline
$p_{8}$	& 		  &			&		  &			&		  &			&		  &			& 4		  & 3		 & 5		& 4		   & 4		  &	3		 & 2		& 2	      \\\hline
$p_{9}$	& 		  &			&		  &			&		  &			&		  &			&		  &	5		 & 4		& 2		   & 2		  &	4		 & 4		& 5	      \\\hline
$p_{10}$& 		  &			&		  &			&		  &			&		  &			&		  &			 & 5		& 4		   & 5		  &	2		 & 3		& 4	      \\\hline
$p_{11}$& 		  &			&		  &			&		  &			&		  &			&		  &			 &			& 4		   & 2		  &	4		 & 3		& 4	      \\\hline
$p_{12}$& 		  &			&		  &			&		  &			&		  &			&		  &			 &			&		   & 2		  &	3		 & 5		& 5	      \\\hline
$p_{13}$& 		  &			&		  &			&		  &			&		  &			&		  &			 &			&		   &		  &	3		 & 4		& 5	      \\\hline
$p_{14}$& 		  &			&		  &			&		  &			&		  &			&		  &			 &			&		   &		  &			 & 4		& 4	      \\\hline
$p_{15}$& 		  &			&		  &			&		  &			&		  &			&		  &			 &			&		   &		  &			 &			& 2	      \\\hline
$p_{16}$& 		  &			&		  &			&		  &			&		  &			&		  &			 &			&		   &		  &			 &			&	      \\\hline
\end{tabular}
}
\end{table}
\begin{table}[ht]
\caption{Comparisons between ALG1 (with $64$ processors) and the epitaxial algorithm}
\centering
{\scriptsize
\label{table:ALG1results}
\begin{tabular}{|r|r|r|r|r|r|r|}
\hline
			&				&	\multicolumn{5}{|c|}{ALG1}																		\\\cline{3-7}
${\it dim}$	&	Epitaxial	&	$T=2$ min.		&	$T=4$ min.		&	$T=6$ min.		& 	$T=8$ min.		&	$T=10$ min.		\\\hline
32			&	55,296		&	55,680			&	{\bf 55,238}	&	{\bf 55,068}	&	{\bf 54,942}	&	{\bf 54,894}	\\
30			&	48,604		&	48,832			&	{\bf 48,576}	&	{\bf 48,428}	&	{\bf 48,418}	&	{\bf 48,392}	\\
28			&	42,676		&	{\bf 42,298}	&	{\bf 42,184}	&	{\bf 42,088}	&	{\bf 42,070}	&	{\bf 42,040}	\\
26			&	36,806		&	{\bf 36,656}	&	{\bf 36,536}	&	{\bf 36,528}	&	{\bf 36,450}	&	{\bf 36,382}	\\
24			&	31,480		&	{\bf 31,074}	&	{\bf 31,004}	&	{\bf 31,004}	&	{\bf 31,004}	&	{\bf 31,004}	\\
22			&	26,608		&	{\bf 26,208}	&	{\bf 26,208}	&	{\bf 26,208} 	&	{\bf 26,208}	&	{\bf 26,208}	\\
20			&	21,956		&	{\bf 21,698}	&	{\bf 21,666}	&	{\bf 21,666}	&	{\bf 21,666}	&	{\bf 21,666}	\\
18			&	17,884		&	{\bf 17,588}	&	{\bf 17,588}	&	{\bf 17,588}	&	{\bf 17,588}	&	{\bf 17,588}	\\
16			&	14,190		&	{\bf 13,916}	&	{\bf 13,916}	&	{\bf 13,916}	&	{\bf 13,916}	&	{\bf 13,916}	\\\hline
\end{tabular}
}
\end{table}
\begin{table}[ht]
\caption{Comparisons between ALG2 (with $64$ processors) and the epitaxial algorithm}
\centering
{\scriptsize
\label{table:ALG2results}
\begin{tabular}{|r|r|r|r|r|r|r|}
\hline
			&				&	\multicolumn{5}{|c|}{ALG2}																		\\\cline{3-7}
${\it dim}$	&	Epitaxial	&	$T=2$ min.		&	$T=4$ min.		&	$T=6$ min.		& 	$T=8$ min.		&	$T=10$ min.		\\\hline
34			&	62,072		&	63,316			&	62,774			&	62,438			&	62,222			&	{\bf 62,026}	\\
32			&	55,296		&	56,060			&	55,468			&	{\bf 55,258}	&	{\bf 55,060}	&	{\bf 54,918}	\\
30			&	48,604		&	48,924			&	{\bf 48,398}	&	{\bf 48,226}	&	{\bf 48,070}	&	{\bf 47,988}	\\
28			&	42,676		&	{\bf 42,382}	&	{\bf 42,138}	&	{\bf 42,006}	&	{\bf 41,946}	&	{\bf 41,894}	\\
26			&	36,806		&	{\bf 36,572}	&	{\bf 36,332}	&	{\bf 36,246}	&	{\bf 36,222}	&	{\bf 36,192}	\\
24			&	31,480		&	{\bf 31,032}	&	{\bf 30,916}	&	{\bf 30,864}	&	{\bf 30,818}	&	{\bf 30,774}	\\
22			&	26,608		&	{\bf 26,084}	&	{\bf 25,922}	&	{\bf 25,860}	&	{\bf 25,850}	&	{\bf 25,842}	\\
20			&	21,956		&	{\bf 21,672}	&	{\bf 21,482}	&	{\bf 21,482}	&	{\bf 21,482}	&	{\bf 21,482}	\\
18			&	17,884		&	{\bf 17,376}	&	{\bf 17,376}	&	{\bf 17,330}	&	{\bf 17,328}	&	{\bf 17,328}	\\
16			&	14,190		&	{\bf 13,730}	&	{\bf 13,730}	&	{\bf 13,730}	&	{\bf 13,730}	&	{\bf 13,730}	\\\hline
\end{tabular}
}
\end{table}
\end{onecolumn}


\begin{thebibliography}{00}
\bibitem{affymetrix1}
\textbf{Affymetrix, Inc.:} {http://www.affymetrix.com}.

\bibitem{fodor1}
Fodor S, Read JL, Pirrung MC, Stryer L, Tsai LA, Solas D:
  \textbf{Light-Directed, Spatially Addressable Parallel Chemical Synthesis}.
  \emph{Science} 1991, \textbf{251}:767--773.

\bibitem{mandoiu1}
Kahng AB, Mandoiu II, Pevzner P, Reda S, Zelikovsky A: \textbf{Border Length
  Minimization in DNA Array Design}. In \emph{Proceedings of the 2nd Workshop
  on Algorithms in Bioinformatics}, Springer LNCS 2002:435--448.

\bibitem{sg1}
Singh-Gasson S, Green RD, Yue Y, Nelson C, Blattner F, Sussman MR, Cerrina F:
  \textbf{Maskless Fabrication of Light-Directed Oligonucleotide Microarrays
  using a Digital Micromirror Array}. \emph{Nature Biotechnology} 1999,
  \textbf{17}:974--978.

\bibitem{invitrogen1}
\textbf{Invitrogen, Inc.:} {http://www.invitrogen.com}.

\bibitem{febit1}
\textbf{Febit, Inc.:} {http://www.febit.com}.

\bibitem{hannenhalli1}
Hannenhalli S, Hubbell E, Lipshutz R, Pevzner PA: \textbf{Combinatorial
  Algorithms for Design of DNA Arrays}. In \emph{Chip Technology}. Edited by
  Hoheisel J, Springer 2002.

\bibitem{mandoiu2}
Kahng AB, Mandoiu II, Pevzner P, Reda S, Zelikovsky A: \textbf{Engineering a
  Scalable Placement Heuristic for DNA Probe Arrays}. In \emph{Proceedings of
  the 7th Annual International Conference on Research in Computational
  Molecular Biology}, ACM 2003:148--156.

\bibitem{mandoiu3}
Kahng AB, Mandoiu II, Reda S, Xu X, Zelikovsky A: \textbf{Evaluation of
  Placement Techniques for DNA Probe Array Layout}. In \emph{Proceedings of the
  IEEE-ACM International Conference on Computer-Aided Design}, IEEE Press
  2003:262--269.

\bibitem{ls1}
Papadimitriou CH, Steiglitz K: \emph{Combinatorial Optimization: Algorithms and
  Complexity}. Dover Publications 1998.

\bibitem{mpi1}
\textbf{The MPI standard:} {http://www-unix.mcs.anl.gov/mpi}.

\bibitem{SABiosciences1}
\textbf{SABiosciences, Inc.:} {http://www.sabiosciences.com/custom.php}.

\end{thebibliography}
\end{document}